\documentclass[twocolumn,showpacs,preprintnumbers,amsmath,amssymb]{revtex4}
\usepackage{amssymb}
\usepackage{color}

\usepackage{graphicx}
\usepackage{dcolumn}
\usepackage{bm}

\begin{document}

\title{Manipulating the momentum state of a condensate by sequences of standing wave pulses}

\author{Wei Xiong}
\author{Xuguang Yue}
\author{Zhongkai Wang}
\author{Xiaoji Zhou} \thanks{Electronic address: xjzhou@pku.edu.cn}
\author{Xuzong Chen} \thanks{Electronic address:
xuzongchen@pku.edu.cn}
\affiliation{School of Electronics Engineering $\&$ Computer
Science, Peking University, Beijing 100871, China}
\date{\today}

\begin{abstract}
We analyze the effects of sequences of standing wave pulses on a
Bose-Einstein condensate (BEC). Experimental observations are in
good agreement with a numerical simulation based on the band
structure theory in the optical lattice. We also demonstrate that a
coherent control method based on such sequences of pulses is very
efficient for experimentally designing specific momentum states.
\end{abstract}

.

\maketitle
\section{Introduction}
Atomic interferometry is very useful in fundamental studies of
coherence, decoherence and phase shifts and for practical precision
measurements, with the example of gravimeters, gyroscopes, and
gradiometers~\cite{Alexander2009rmp}. Bose-Einstein Condensate (BEC)
based atomic interferometry provides high contrasts, long
integration times and the possible use of small
devices~\cite{Yoshio2000pra, R. E. Sapiro2009pra}. In such atomic
interferometry, coherent momentum manipulation is very effective for
splitting and recombining the condensate~\cite{Kozuma1999PRL,
Deng1999PRL, Ovchinnikov1999PRL}, hence realizing the interference.
In some precision measurements, the accumulated phase is positively
correlated to the atomic velocity, so that the larger the atomic
momentum is, the more precise the measurements can be if the
measuring time is unchanged.

The diffraction of atoms from standing wave light, which is usually
divided into three regimes as Bragg, Raman-Nath and channeling
regimes~\cite{Keller1999apb} by the interaction intensity and
duration, is one of the common methods to coherently manipulate the
momentum states of a condensate~\cite{GuptaLeanhardt2001,
Ptitchard2001, Meystre2001book, Wu2005pra}. One-pulse Bragg or
Raman-Nath scattering can be applied for momentum states'
preparation, but those techniques are then limited by constraints on
the pulse properties~\cite{Keller1999apb}. A momentum manipulation
method by multi-pulse standing waves is proposed in
~\cite{Edwards2010pra}, where the momentum states can be designed,
but not observed yet, and the pulses are still restricted in the
Raman-Nath regime.

In this paper, we apply a method for flexible manipulation of the
atomic momentum states, where the standing wave pulses are less
limited in pulse intensities and durations. The atomic diffraction
from one, two, three and four standing wave pulses are demonstrated
in our experiments and systematically analyzed by the band structure
theory of one-dimension optical lattice. With this method, we are
able to design and realize several specific momentum states, which
may be applied in atomic interferometry. In principle, this method
could be used for designing a wide range of possible target states.

This paper is organized as follows. In Section II, a theory to
interpret the scattering process by a sequence of pulses is
presented, where the standing wave is treated as a one-dimensional
optical lattice. We derive a concise expression for calculating the
probability of each momentum state at the end of the process. In
Section III, the experiments of one, two, three and four pulses'
scattering are demonstrated and compared to theoretical simulations.
It is found that a correction due to momentum dispersion may be
introduced into the theory for a better agreement with the
experimental results. Section IV presents the experimental
realization of several useful momentum states by coherent control.
Section V contains the discussions and the conclusions.

\section{Theoretical model}
We first consider a non-interacting condensate being diffracted by a
sequence of square shaped standing wave pulses with the successive
durations $\tau_i (i=1,2,...,s+1)$, separated by the intervals
$\tau_{fi} (i=1,2,...,s)$. The standing wave consists of a pair of
laser beams far-detuned enough to suppress the spontaneous emission.

The periodic potential (one-dimension optical
lattice)~\cite{Morsch2006,Denschlag2002jpb} introduced by the ac
Stark shift can be described as $V(x)=U_0\cos^2(k_Lx)$, with the
trap depth $U_0$ and the laser's wave vector $k_L = 2\pi /
\lambda_L$ ($\lambda_L$ is the wavelength of the laser). The lattice
leads to a band structure of the energy spectra, of which the
eigenvalues of the energy $E_{n,q}$ and eigenvectors $|n, q\rangle$
(Bloch states) are labeled by the quasi-momentum $q$ and the band
index $n$, and they satisfy the equation:
\begin{equation}\label{eig}
  \hat H\left| {n,q} \right\rangle  = {E_{n,q}}\left| {n,q} \right\rangle
\end{equation}
where the Hamiltonian $\hat H = {{\hat p}^2}/2M + {U_0}{\cos
^2}({k_L}x)$, with the atomic momentum $\hat p$ and the atomic mass
$M$. The Bloch states form a quasi-momentum space. In the lattice,
the spatial periodicity of the wave function results in separated
peaks in momentum space, corresponding to the reciprocal lattice
vector $2k_L$.

When a condensate with an initial momentum $p_{m_0}=\hbar(q + 2m_0
k_L)$ ($\hbar$ is the Plank constant, $-1 \le q \le 1$ , ${m_0} =
..., - 1,0,1,...$) is abruptly loaded into a lattice, the wave
packet can be described as a superposition of the Bloch states:
\begin{equation}\label{proj}
  |\Psi(t=0)\rangle=\sum_{n=0}^\infty|n,q\rangle\langle n,q|p_{m_0}\rangle
\end{equation}
where $\langle n,q|p_{m_0}\rangle = c_{n,q}(m_0)$. The $n^{th}$
Bloch state evolves independently as $e^{-iE_{n,q}t/\hbar}$, and the
total wave function evolves as
\begin{equation}\label{evolve}
  |\Psi(t)\rangle=\sum_{n=0}^\infty c_{n,q}(m_0)e^{-iE_{n,q}t/\hbar}|n,q\rangle
\end{equation}
While the incident light is switched off after the duration
$\tau_1$, the wave function is projected back to the momentum space
from the quasi-momentum space. The coefficient $b(m_0,m,\tau_1)$ of
each $|p_m\rangle$ state ($m = ..., - 1,0,1,...$) can be acquainted
as:
\begin{equation}\label{firstPulse}
  b(m_0,m,\tau_1)=\sum_{n=0}^\infty c_{n,q}(m_0)c_{n,q}(m) e^{-iE_{n,q}\tau_1/\hbar}
\end{equation}
For a zero initial momentum of the condensate, the subscript $q$ can
be omitted for simplification and $m_0=0$. For one pulse scattering,
the population of the $|p_m\rangle$ state is
$P^{(1)}_m=|b(0,m,\tau_1)|^2$. It can be seen that the probabilities
of the momentum states after one scattering pulse depend on the
lattice depth and the pulse duration. The lattice depth determines
the band structure and is reflected in the terms $c_{n,q}$. The
pulse duration influences the phase evolution of each Bloch state as
$e^{-iE_{n,q}\tau_1/\hbar}$.

The multi-pulse process, which consists of a number of single pulses
and intervals can be solved as follows. The wave function of the
condensate after the first pulse $\tau_1$ can be derived from
Eq.~(\ref{firstPulse}) as
\begin{equation}
  |\Psi(\tau_1,t)\rangle=\sum_m b(m_0,m,\tau_1)e^{-iE^{(m)}t/\hbar}|2m\hbar k_L\rangle
\end{equation}
After the first interval $\tau_{f1}$ and the second pulse $\tau_2$,
the population of the $|p_{m}\rangle$ state can be achieved as
\begin{equation}\label{twopulse}
\begin{split}
  P^{(2)}_m=\left|\sum_{m_1}b(m_0,m_1,\tau_1) e^{-iE^{(m_1)}\tau_{f1}/\hbar}
  b(m_1,m,\tau_2)\right|^2
\end{split}
\end{equation}
As shown in Eq.~(\ref{twopulse}),the population is affected by the
two pulses, the first one corresponding to $b(m_0,m_1,\tau_1)$, and
the second one corresponding to $b(m_1,m,\tau_2)$. During the
interval $\tau_{f1}$, the phase of the $|p_{m_1}\rangle$ state
evolves along the time as $e^{-iE^{(m_1)}t/\hbar}$, where
${E^{(m_1)}} = {(2m_1\hbar {k_L})^2}/2M = 4{m_1^2}{E_R}$ is the
kinetic energy, and ${E_R} = {(\hbar {k_L})^2}/2M$ is the single
photon recoil energy. The interval $\tau_{f1}$ produces a phase
shift $e^{-iE^{(m)}\tau_{f1}/\hbar}$ and contributes to the momentum
distribution.

In the same way, the population of the $|p_{m}\rangle$ state after
$(s+1)$ pulses can be achieved as:
\begin{equation}\label{sPulses}
\begin{split}
  P^{(s+1)}_m=\left|\sum_{m_1,m_2,\cdots,m_s}
  \prod_{i=1}^{s+1} b(m_{i-1},m_{i},\tau_i)
  \prod_{i=1}^{s}e^{-iE^{(m_i)}\tau_{f_i}/\hbar} \right|^2
\end{split}
\end{equation}
with $m_0=0$, and $m_{s+1}=m$.

From the analysis above, the momentum distribution after a sequence
of pulses' scattering is influenced by not only the lattice pulses
with the term $\prod_{i=1}^{s+1} b(m_{i-1},m_{i},\tau_i)$, but also
the intervals among the pulses as reflected in the term
$\prod_{i=1}^{s}e^{-iE^{(m_i)}\tau_{f_i}/\hbar}$. Although the
populations of the momentum states do not change during the
intervals, the phase-evolution rates of the momentum states with
different kinetic energies are not identical. The phase deviations
between the states oscillate from 0 to $2\pi$ with the interval, and
the heterogeneously accumulated phases change the distribution of
the condensate in the quasi-momentum space.

\section{Experiments of standing wave pulse sequences}
\begin{figure}
\centering
\includegraphics[width=8.5cm]{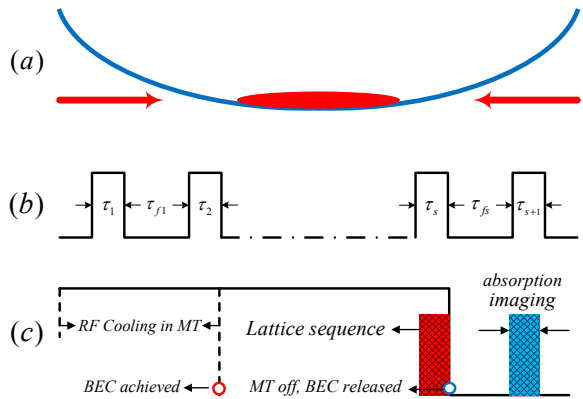}
\caption{(a) A pair of counter-propagating light beams are applied
to a condensate in magnetic trap. (b) The scattering process
consists of a sequence of standing wave pulses, which contained $s$
intervals with widths $\tau_{fi}$ (i=1,2,...$s$) and $s+1$ light
pulses with durations $\tau_i$ (i=1,2,...$s+1$). The incident
light's wavelength is $852~nm$ and its maximum intensity can reach
$120E_R$. (c) The procedure for the experiments is shown. The
condensate is exposed to a sequence of standing wave pulses and then
released from the magnetic trap. The absorption images of the
condensate can be observed after the free falling. } \label{exp}
\end{figure}

We performed the experiments of a condensate in a magnetic trap (MT)
(see Fig.~\ref{exp}(a)) being scattered by a sequence of standing
wave pulses (see Fig.~\ref{exp}(b)). As shown in Fig.~\ref{exp}(c),
after pre-cooling, a cigar shaped $^{87}$Rb condensate of
$2\times10^5$ atoms in $5{}^2{S_{1/2}}\left| {F = 2,{M_F} = 2}
\right\rangle$ state was achieved by the radio frequency (RF)
cooling in the magnetic trap, of which the axial frequency is
$20$~Hz and the radial frequency is $220$~Hz~\cite{Yang2008pra,
Zhou2010pra}. A pair of counter-propagating laser beams, of which
the durations were controlled by an acousto-optical modulator, and
the amplitudes were adjusted by the injection current of a
tapered-amplifier, were applied to the condensate along the axial
direction. The linear polarized incident light at the wavelength
$\lambda_L=852$~nm was focused with a waist of $110~\mu$m to cover
the condensate. The trap depth, which was calibrated by
Kapitza-Dirac scattering experimentally, reached $120E_R$,
corresponding to the light power of $320$~mW. The incident light and
the magnetic trap were simultaneously shut after the BEC-light
interaction. After $30\ ms$ free falling and ballistic expansion,
the momentum distribution of the condensate was pictured by
absorption imaging.

The lattice in our experiments is quite deep, so we concentrate on
the short-pulse diffractions to avoid the de-coherence and heating
effects of long pulses relevant for Bragg scattering. However, for
more flexible momentum manipulation, our pulses are not so short as
the Raman-Nath pulses~\cite{Huckans2009arXiv} used in previous
works.

A brief instruction to the Raman-Nath regime is given in the
following for comparison. In the scattering process, the evolution
during the free evolution intervals is analyzed as in previous
section, while the effect of lattice with adequately short duration
$\tau$ can be analytically solved by the $Schr\ddot odinger$
equation $i\partial \left| {\Psi (t)} \right\rangle /\partial t =
\hat H\left| {\Psi (t)} \right\rangle$, after omitting the atomic
kinetic energy term ${{\hat p}^2}/2M$ in the Hamiltonian. This
approximation can be made while the displacement of the scattered
atoms during the interaction time is much smaller than the spatial
period of the standing wave. Equivalently, the standing wave
duration $\tau$ and the single photon recoil frequency ${\omega _r}
= {\hbar k_L^2}/2M$ have to fulfill $\tau \ll 1/{\omega _r}$. The
pulse is able to split a stationary condensate into components with
symmetrical momenta $p_n=2\hbar k_L (n=0,\pm1,\pm2, ...)$, with
corresponding populations $P_n=J^2_n(U_0\tau/2\hbar)$, where
$J_n(z)$ are Bessel functions of the first kind.

\begin{figure}
\includegraphics[width=8.5cm]{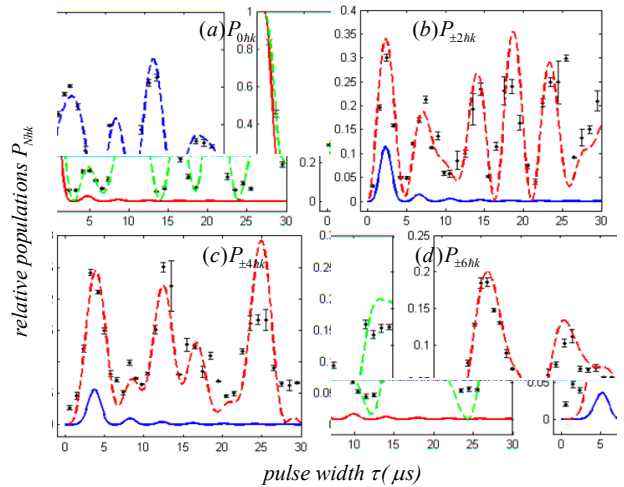}
\caption{Single pulse scattering of condensate: The black dots
represent the experiment results. The blue solid line is the
theoretical analysis with Raman-Nath approximation. The red dashed
line is the numerical simulation with band structure theory. Figure
(a), (b),(c) and (d) correspond to the relative populations of the
condensates with the momenta $0\hbar k$, $\pm 2\hbar k$, $\pm 4\hbar
k$ and $\pm 6\hbar k$ respectively.} \label{onePulse}
\end{figure}

First we demonstrate a one-pulse scattering experiment. A condensate
is exposed to a standing wave pulse with depth $100E_R$ and duration
varying from 0 to 30$\mu s$. The relative populations of the
condensates with the momenta $0\hbar k$, $\pm 2\hbar k$, $\pm 4\hbar
k$ and $\pm 6\hbar k$, corresponding to Fig.~\ref{onePulse}(a), (b),
(c) and (d) respectively, are measured and theoretically analyzed.
In addition, the theoretical analysis with the Raman-Nath
approximation, is also shown in the figure for comparison. It can be
seen that within $3\mu s$ the theoretical analysis with the
Raman-Nath approximation (blue solid line) is close to the
experimental results (black dots), and so is the theoretical
analysis with band structure theory (red dashed line). When the
pulse duration exceeds $3\mu s$, the analysis with the Raman-Nath
approximation gradually goes far away from the experimental results,
while the numerical simulation with band structure theory still
agrees with the experimental results along the whole time scale. As
shown in Fig.~\ref{onePulse}, the probability of each momentum state
oscillates with the pulse duration as described by the band
structure theory. It is clear that, in the single pulse scattering
process, the band structure theory works well not only for the short
pulse but also for the longer pulse, because the atomic motion has
been taken into account. So the atomic diffraction by a single
standing wave pulse can be predicted in a wider range of pulse
duration with the band structure theory.

Then we increase the number of pulses in the experiments to explore
the extra factors influencing the momentum distributions. Two groups
of experiments are carried out, where one consists in two two-pulse
sequences and the other uses a train of three pulses or a train of
four pulses. In every sequence, all the pulses are the same and all
the intervals are identical to make the experiments more convenient
to carry out. For further comparison between the band structure
theory and the analysis in the Raman-Nath regime, every single pulse
is made short enough for the Raman-Nath approximation.

\begin{figure}[t]
  \centering
  \includegraphics[width=8.5cm]{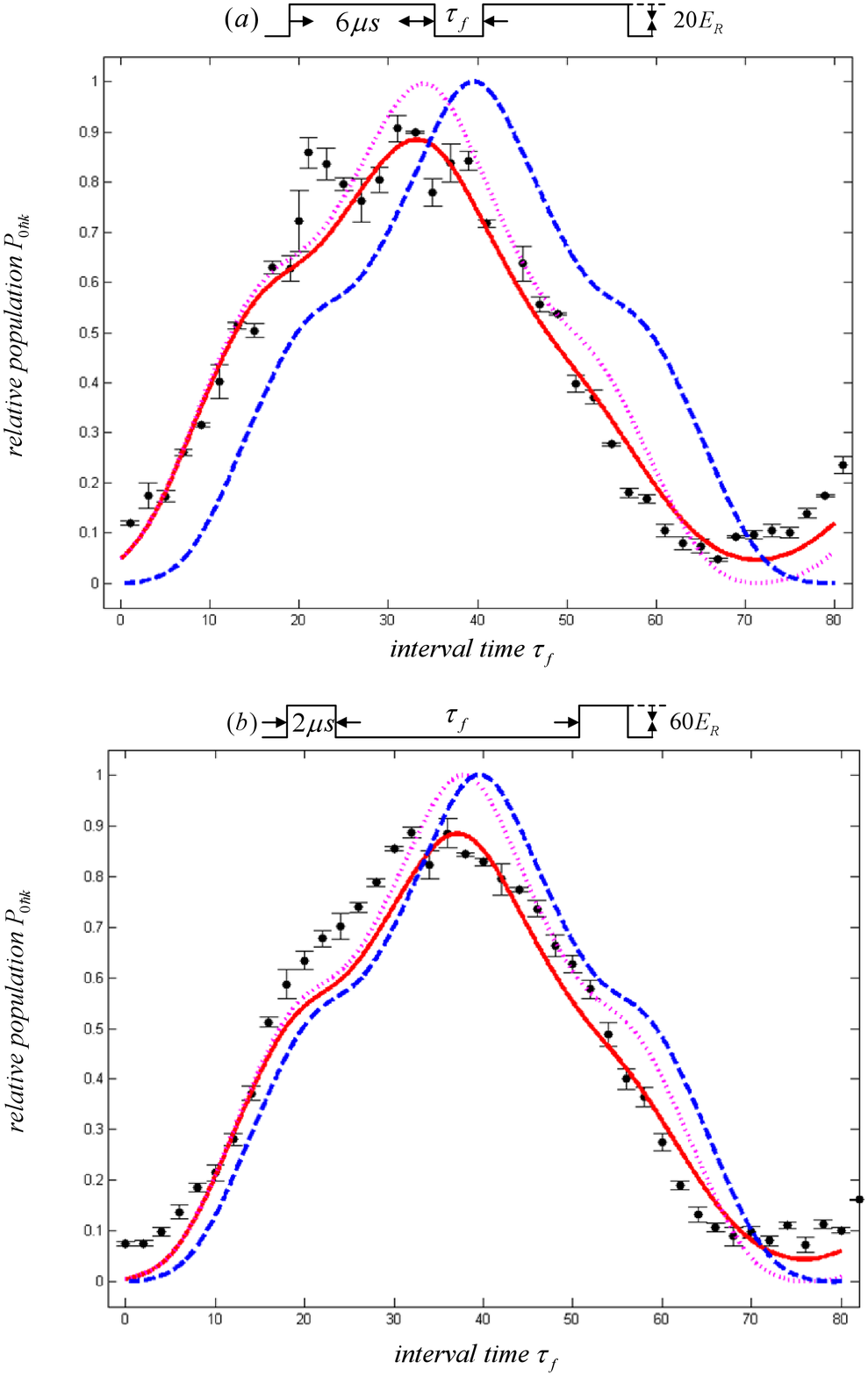}
\caption{Two-pulse scattering of the condensate: The relative
populations of the stationary condensate $P_{0\hbar k}$ versus the
varied intervals $\tau$ is shown. The parameters of the experiments
are described above each figure. The black dots are the experimental
results. The blue dashed line is the analysis with Raman-Nath
approximation. The magenta dotted line is the numerical simulation
with band structure theory. The red solid line is a numerical
simulation taking into account the momentum dispersion.} \label{two}
\end{figure}

Two experiments of two-pulse scattering are demonstrated in
Fig.~\ref{two}, in which the relative populations of the stationary
condensate $P_{0\hbar k}$ versus the varied intervals $\tau$ is
shown. The parameters of the scattering pulses used in different
sequences are chosen to be of the same products of the lattice depth
and the pulse duration, so that each pulse affects the condensate
equivalently. As shown in the figure, the intervals actually affect
the final momentum distribution, and the theoretical analysis with
the band structure theory and Raman-Nath approximation both picture
well the evolution of the atomic distributions versus the interval
between the two pulses. The results of two-pulse scattering can be
explained as the fact that since the phase shift accumulated during
the interval varies harmonically from $0$ to $2\pi$, the probability
of the stationary condensate oscillates between the minimum and the
maximum. When the phase shift is $2\pi$ with the interval $\pi \hbar
/2{E_R}$ (around $80\mu s$), the wave function is little affected by
the interval and the two pulses diffract the condensate as one
combined pulse to make the probability $P_{0\hbar k}$ the minimum.
While the phase shift is $\pi$ with the interval $\pi \hbar /4{E_R}$
(about $40\mu s$), the second pulse produces an effect opposite of
the first one and diffract the non-stationary components of the
condensate back to the stationary one and make the probability
$P_{0\hbar k}$ the maximum.

It also can be acquainted from Fig.~\ref{two} that the numerical
simulation with band structure theory is much closer to the
experimental results than the analytical solution with Raman-Nath
approximation. It is conjectured that the phase evolution during the
scattering process makes the difference. The phase evolution in the
scattering process is neglected in the Raman-Nath approximation, but
not in the numerical simulation with band structure theory. Although
the duration of the scattering process is short, the phase shifts in
the scattering process still increase. The phase shift in the
scattering process needs to be taken into account and influences the
final momentum distribution. As a result, the longer the scattering
pulse is, the larger the difference is. Although the maximum of the
probability $P_{0\hbar k}$ corresponds to the interval $\pi \hbar
/4{E_R}$, the two-pulse experiments in Fig.~\ref{two} can clearly
show that the longer pulse leads to the larger difference. In
Fig.~\ref{two} (a), the pulse duration is $6\mu s$, the probability
$P_{0\hbar k}$ reaches the maximum with the interval $34\mu s$. In
Fig.~\ref{two} (b), the pulse duration is $2\mu s$, the probability
$P_{0\hbar k}$ gets to the top with the interval $38\mu s$.

Nevertheless, there is still some obvious deviation between the
simulation and the experimental results. It is observed that the
momentum width has been expanded after the former pulse, because of
the $s$-wave scattering between the different momentum states.
Consequently, this dispersion process is approximated to an initial
momentum width of $\sim0.1\hbar k_L$ on average to optimize the
numerical simulation. Unlike the analysis without momentum width,
phase evolution is different for different initial momenta and
results in a phase dispersion. The quasi modes obtained at the end
of the diffraction process result from the linear superposition of
final states obtained after time evolutions of the different momenta
populated the initial BEC. It can be seen from Fig.~\ref{two} that
the approximation is effective.

As discussed in ~\cite{Li2008prl}, the maximum of the probability
$P_{0\hbar k}$ will never reach $1$ thanks to the imperfect optical
lattice. In our case, the momentum expansion is an explanation of
the similar situation as shown in Fig.~\ref{two}. Since the momentum
width is considered, the phase shift is populated around $\pi$ with
a width, instead of a definite $\pi$, with the interval $\pi \hbar
/4{E_R}$. In other words, there is no interval that accumulates a
phase shift exactly equal to $\pi$, so with any interval, the second
standing wave pulse is not able to diffract all the condensates back
to the stationary part.

\begin{figure}[t]
  \centering
  \includegraphics[width=8.5cm]{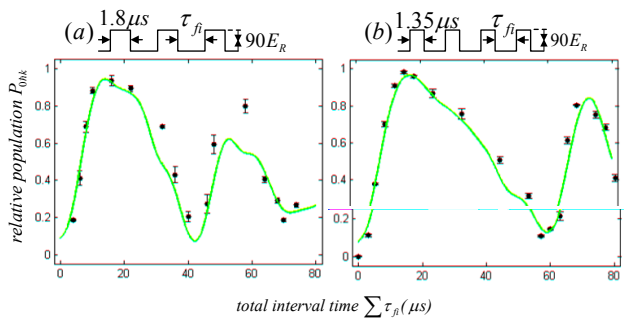}
\caption{Three-pulse and four-pulse scattering of the condensate:
The relative populations of the stationary condensate $P_{0\hbar k}$
versus the varied total intervals ${\sum \tau  _{fi}}$ is shown. The
parameters of the experiments are described above each figure. The
black dots are the experimental results. The red solid line is a
numerical simulation taking into account the momentum dispersion.}
\label{multi}
\end{figure}

The experiments of one three-pulse scattering and one four-pulse
scattering are demonstrated in Fig.~\ref{multi}, where the relative
populations of the stationary condensate $P_{0\hbar k}$ versus the
varied total intervals ${\sum \tau_{fi}}$ is shown. In each
experiment, the total interaction intensity $\sum {{U_0}\tau }$ is
the same, with different number of the pulses. These two experiments
show that even though the total interactions and intervals are the
same, the different processes of phase accumulations in the two
kinds of pulse sequences result in distinct momentum distributions.
We directly apply the band structure theory with momentum dispersion
to analyze the experiments in Fig.~\ref{multi}, and the corrected
simulations agree with the experiments quite well.

\section{Manipulate the momentum states as design}
\begin{figure}
\centering
\includegraphics[width=8.5cm]{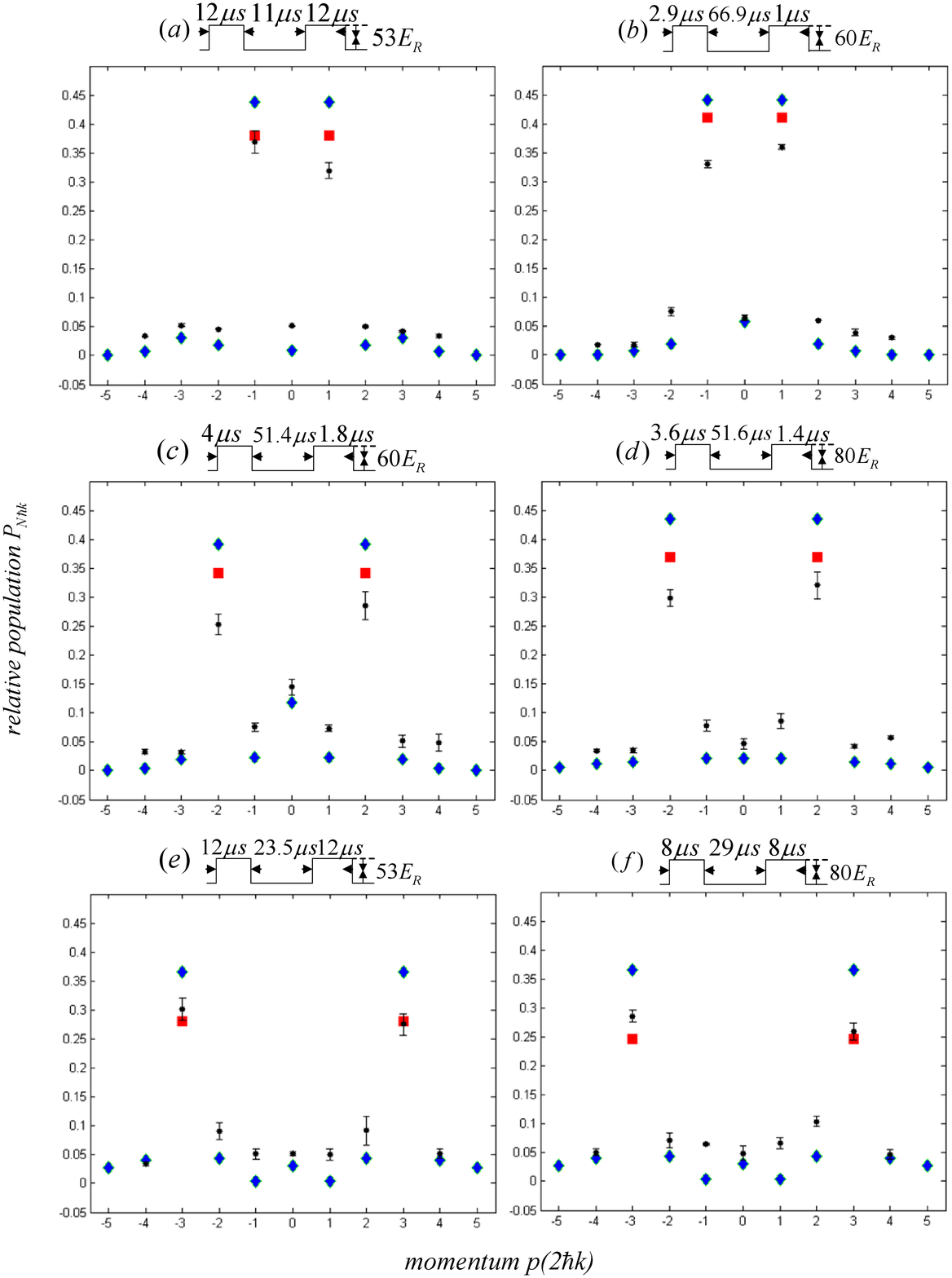}
\caption{Experimental realization of designed momentum states. The
expected momentum state is $\pm2\hbar k_L$((a) and (b)), $\pm4\hbar
k_L$((c) and (d)) and $\pm6\hbar k_L$((e) and (f)). The pulse
sequences are shown above each figure. The black round dots are
experimental results. The blue diamond dots are the expectations
based on the design. The red square dots are the modified design
with momentum width, which agree with the experiments better.}
\label{design}
\end{figure}

The experiments and the numerical simulations above have shown the
possibility and feasibility of the manipulation of a condensate's
momentum states. We manage to design several two-pulse sequences to
achieve high contrast momentum states such as $\left| { \pm 2\hbar
k} \right\rangle$, $\left| { \pm 4\hbar k} \right\rangle$ and
$\left| { \pm 6\hbar k} \right\rangle$, which may be useful in
atomic interferometry~\cite{Beattie2009pra, Rohwedder2001epjd}. For
each state, we apply two totally different two-pulse sequences to
show the flexibility of the method. The general method to achieve
the target states is to find out the condition of the minimum of the
square deviation ${\Delta ^2} = \sum\limits_{m =  - \infty }^{ +
\infty } {{{(P_m^g - P_m)}^2}}$, where ${P_m^g}$ is the probability
of $\left| {2m\hbar {k_L}} \right\rangle$ in the goal state, and
${P_m}$ is that generated by the sequence. A second method, as the
target is to obtain the highest population of some certain momentum
state, consists in scanning the whole set of initial conditions and
choose the one corresponding to the maximum value of the desired
population. We apply the two methods above separately and obtain the
same pulse sequences. As shown in Fig.~\ref{design}, the
experimental results (the black round dots) agree well with the
expectations of the designs (the blue diamond dots), whether the
pulses are in the Raman-Nath regime (see Fig.~\ref{design}(b)) or
not (see others in Fig.~\ref{design}). When the momentum dispersion
is being considered, the expected momentum distributions (the red
square dots) get closer to the experiments, where the figures only
display the relative populations of the target states and omit the
others for the figures being more clear.

It can be seen from Fig.~\ref{design} that the momentum width
correction can improve the precision of the prediction with our
method. The average relative deviation between the experimental
results and the expected values without including the momentum width
is $25.03\%$, while the deviation is decreased to $13.15\%$ with the
correction.

An asymmetry of the momenta can be observed in Fig.~\ref{design},
and it may ascribe to the following factors. Besides the measurement
error, there is an imperfection of the standing wave, brought forth
by the unbalanced intensity of the laser beams. External field (such
as the magnetic trap) fluctuations during the scattering process may
also affect the momentum distribution.

\section{Discussion and Conclusion}
The band structure theory is a global method to deal with the
standing wave scattering a condensate, while Bragg and Raman-Nath
scattering are two special situations which can be analytically
solved with their respective approximations. In the Bragg regime,
the potential height introduced by the standing wave is restrained
below $4E_R$ and that leads to the difficulty of generating higher
order momentum states. In the Raman-Nath regime, the intensity of
the standing wave is not limited so that higher order momentum
states can be generated symmetrically~\cite{Sapiro2009njp,
Gadway2009OL}. However the pulse duration has to be short enough to
neglect the atomic motion, so the momentum states can not be
predicted in this regime if the pulse duration is slightly longer.
In our work, the scattering can be well explained and numerically
analyzed within a much wider range of pulse intensity and duration.
So it is natural that more momentum states can be generated.


In our article, we compared the scattering by one pulse and that by
a train of pulses. Some valuable states, such as $\left| { \pm
2\hbar k} \right\rangle$, $\left| { \pm 4\hbar k} \right\rangle$ and
$\left| { \pm 6\hbar k} \right\rangle$ states with high contrast,
can not be realized by the single pulse scattering, while they can
be realized by a sequence of standing wave pulses. A sequence of
lattice pulses is a more effective and flexible tool for momentum
manipulation. It can generate much more useful momentum states than
the ones demonstrated in our work. In the future, more parameters
could be changed for the design to obtain better results.

Even though the numerical simulation is corrected to take into
account the momentum dispersion, some deviations between the
experiments and the simulation still exist. The inaccuracy of the
lattice-depth calibration, which is $5\%$ at least, is one of the
reasons. The phase shift introduced by the magnetic trap is another
one, while the influence is estimated to be within $0.03\%$, which
is below the experimental uncertainty. The heating and momenta
exchange during the s-wave scattering may also lead to some
differences, which need further study.

In conclusion, we developed a method for more flexible manipulation
of the condensate's momentum states, where the momentum states can
be controlled by standing wave pulses in a wider range of pulse
intensity, duration or quantity. The experiments of a condensate
being scattered by a sequence of standing wave pulses are
demonstrated. A global theory, treating the standing wave as an
optical lattice, is applied to explain the experiments. With this
theory, we are able to design pulses sequences for realizing states
such as $\left| { \pm 2\hbar k} \right\rangle$, $\left| { \pm 4\hbar
k} \right\rangle$ and $\left| { \pm 6\hbar k} \right\rangle$, and
experimentally realize them, which may be applied in atomic
interferometry to improve the measurement precision.

\section{Acknowledgement}
We would like to thank Thibault Vogt for critical reading of our paper. This work
is supported by the National Fundamental Research Program of China
under Grant No. 2011CB921501, the National Natural Science
Foundation of China under Grant No. 61027016, No.61078026,
No.10874008 and No.10934010.

\end{document}